# Random Number Hardware Generator Using Geiger-Mode Avalanche Photo Detector

D. Beznosko [1], T. Beremkulov, A. Duspayev, A. Iakovlev, A. Tailakov, M. Yessenov

Nazarbayev University Cosmic Rays and Particles group, NU Physics Department, 53 Kabanbai Batyr ave, Astana, KZ 010000

## Abstract

This paper presents the physical concept and test results of sample data of the high-speed hardware true random number generator design based on typically used for High Energy Physics hardware. Main features of this concept are the high speed of the true random numbers generation (tens of Mbt/s), miniature size and estimated lower production cost. This allows the use of such a device not only in large companies and government offices but for the end-user data cryptography, in classrooms, in scientific Monte-Carlo simulations, computer games and any other place where large number of true random numbers is required. The physics of the operations principle of using a Geiger-mode avalanche photo detector is discussed and the high quality of the data collected is demonstrated.

## Introduction

      High Energy Physics (HEP) in many ways has influenced the progress, creating new technologies that are later adapted for everyday use. Here, we propose the use of the available in HEP for the last decade miniature photo sensors as a basis for the hardware-based random number generator that could be commercialized in the future. This project was carried out as part of the research program by the Nazarbayev University Cosmic Rays and Particles group [1].

      The hardware produced random numbers are used by large companies (such as banks, communications and cell phone companies) and by many countries' government planning offices in their simulations of the economy growth and similar tasks. Due to the low speed, low availability and/or prohibitively high cost of such devices, analytical algorithms are used instead in small companies, in science or by the end-users. The downside of any analytical (e.g. software) method is that it is not truly random, thus presenting a weakness that can be exploited for malicious purpose. Therefore, a simple, robust and affordable solution is necessary.

      The physics principle of the device involves the detection of the single photons incident onto the Geiger-mode avalanche photo detector [2], which is a true random process. These detectors have been developed recently and are available from several manufacturers, the sensors from two of these are tested in [3] and in [4]. They have been already used in large scale High Energy physics experiments, e.g. in the T2K-ND280 pi-zero detector [5] usually in conjunction with plastic scintillator [6]. The main features for these devices are

---

[1] Email: dmitriy.beznosko@nu.edu.kz

high gain (~$10^6$), robustness, low biasing voltage, high sensitivity in optical range (~25%) and relatively low cost. Future experiments considering their wide-scale use are Horizon-T [7] [8] and HT-KZ [9] cosmic rays detector systems.

## Experimental Setup

The 400-pixel square Hamamatsu [10] MPPC photodiode [11] was used for this setup. With the biasing voltage of ~-70.10V, it provides the gain of ~$7 \cdot 10^5$ and the detection efficiency of up to 30% at 500nm. A Bivar [12] SM1204PGC Light Emitting Diode (LED) in the metal box with a small opening was used as a signal source, and a CAEN [13] DT5743 digitizer was used to record the signal. The schematic of the setup is shown in Figure 1. Since the pulse width from the MPPC is couple of tens of ns, the system can be run at several MHz (thus collecting Mbits of data per second), but we were limited by DAQ system to about 1 kHz.

For the low cost solution, it is possible to realize the whole setup on a single controller chip with LED and MPPC connected to it directly, thus making the possible cost of such a device much lower than comparable ones on the market. The clear separation of signal-no signal detection for MPPC (and similar devices) allows for a digital readout of the signal, thus eliminating the most costly component of the setup – ADC. With a simple comparator, the controller can read digital output from the MPPC directly.

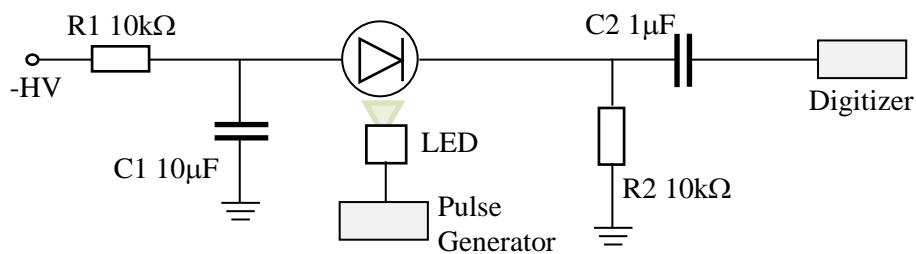

**Figure 1: Random generator Setup Schematics.**

## Experimental Procedure

The experimental basis of the idea uses the fact that the MPPC photodiode has a clear photoelectron (PE) separation both in amplitude and in charge. The light from the LED via a small opening in the metal housing is incident onto the photodiode producing the signal output shown in Figure 2 for both the amplitude (left) and area (right). The rightmost peak is the pedestal (that is, zero photons detector), the next peak to the left is the 1$^{st}$ PE, next is the 2$^{nd}$, etc. since the signal amplitudes are negative. As a simple comparator will work with the amplitudes rather than areas, only amplitude will be used further to emulate it. Note that the signal powering the LED is also the trigger for the data collection in order to reduce the probability of the external noise to enter the signal. The detection gate width used was 100ns and was centered on the MPPC average response position. The area is the algebraic sum of all amplitudes within the gate.

As the area of the pedestal in Figure 2 (left) has approximately 50% of the events in it, the threshold was chosen to be at -0.008V. No extra care was taken for the pedestal to be 50% exactly or the area to be stable long-term due to the analysis method used. Both the light yield of the LED and the detection efficiency of the MPPC are weekly dependent upon the temperature and photodiode bias [14], both are stable on a small time scale of several tens of seconds, which is the timescale of the data collection runs (clarified in next chapter); thus, we assume the stability of the conditions. The biasing voltage was within ±0.002V and changes of temperature were not detectable.

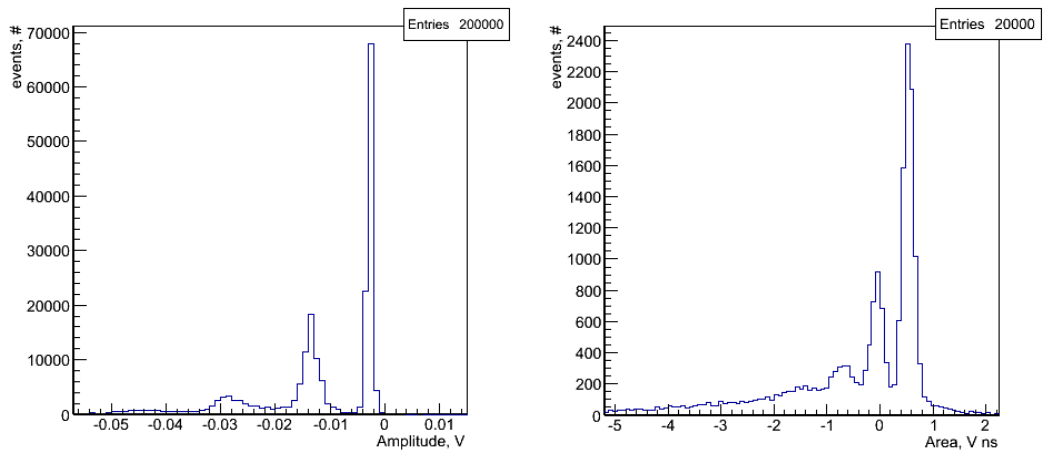

**Figure 2: MPPC photoelectron signal separation in amplitude (left) and charge (right).**

## Theoretical Considerations

Let us consider the statistics of the light detection process. The amount of photons that falls onto the photodiode follows the Poisson distribution. The light detection efficiency of the photodiode is constant (changing very weakly with applied bias value) and is on the order of ~30%. With the light intensity at low light level of few photons per pulse, the photodiode sometimes detects no photons at all, sometimes one or more. This is evident from Figure 2, where the areas (right) and amplitudes (left) of each peak follow the distribution. However, if we set a threshold (at ~ -0.005V) that separates the pedestal from the signal, the resultant is the integral probability of signal being detected or not, and is now linear. The light intensity of the source could be adjusted to have this probability at approximately 50%.

As the data is acquired, the values below it will be converted into the bit of value 1, and above it to a bit with a value of 0. The resultant output file is the binary data written as a sequence of 1s and 0s without the separation into bytes or words, e.g. a stream of bits. The separation is done as the second stage, when the data is being analyzed.

Before any test is performed, one needs to note that the resultant data, while being random, does not exactly contain even amounts of ones and zeros. The reason for that is the extreme difficulty to constrain all physical conditions of the system to keep the generator producing the exact 50-50 output all the time. However, there is a simple method to even out the data without losing the randomness of it called the AMLS [15], it is a randomness extraction algorithm with the code implementation of it taken from [16]. The code output is the bits sequence with the same amount of 0s and 1s, but it requires an initial source of randomness to function.

The combination of the MPPC+LED presents to be such a randomness source and the AMLS code allows this simple generator to produce high quality data. This is also a reason why large-scale changes in temperature and other operating conditions do not affect the output – the AMLS code is run every (few) second(s) on the data sample collected. Then next data sample is now independent from the previous, etc. thus insuring stable operations.

# Testing of Generated Random Numbers

A vast number of testing suits are available to check that the given sequence of numbers is truly random. For each of them, a theoretical result is also known for a sample of 'perfect' random data, thus allowing a comparison. As this paper is more about the generator design using the MPPC sensor, we will present here the results of the most illustrative tests that indicate the quality of randomness of the data collected.

The simple visual test is to graphically represent the data. The bits are read by 8 as a single unsigned integer and the resultant value (0-255) is plotted as a pixel brightness (0 being black) on a grayscale image in Figure 3. The patterns on the image will show the presence of the non-randomness. This test, while almost primitive, yields good results in detecting large-scale repeatability or patterns in the data.

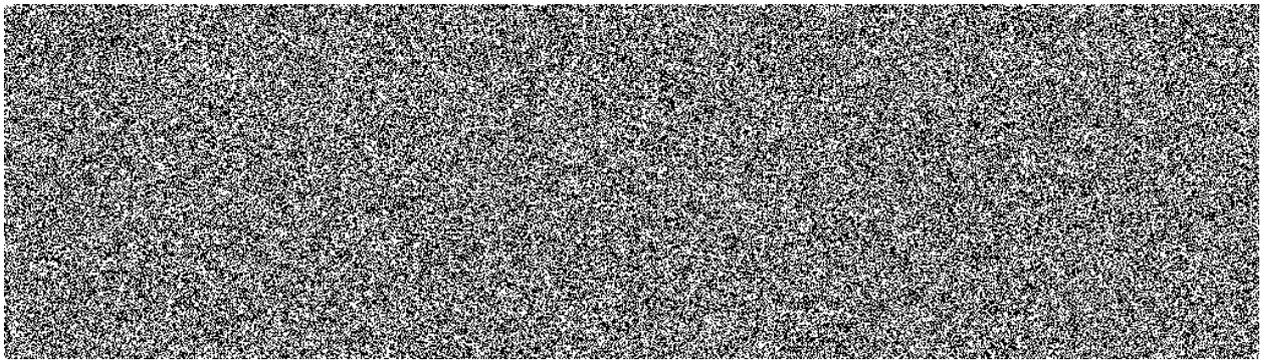

**Figure 3: Random data as an image.**

Another graphical test is to read the data as 16bit signed integers and plot them as a histogram (Figure 4). Here, the possible values are from 32767 to -32767. Two aspects of the data are tested – mean of the data (for ideal data, expected to be 0 deviating randomly from this value), and the slope of the fitted line $p1*x+p0$, where the expectation is $p1=0$ for truly random data. The plot and the fit are done using ROOT [17]. From the figure, we see that the actual values for the mean and the slope are close to expected values showing absence of any systematic bias in the data.

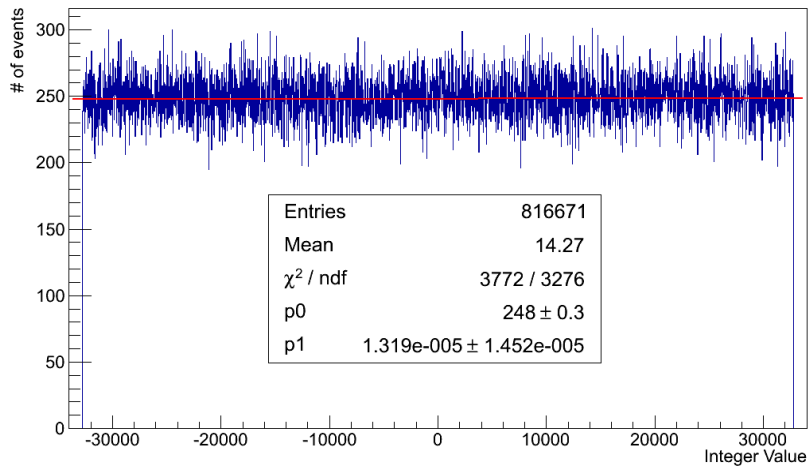

**Figure 4: Random data as a histogram with a fit.**

*A Pseudorandom Number Sequence Test using ENT*

ENT program [18] includes different random number tests. More detailed information about this program and tests that it implements on data file may be found on the official web-site of the program. Above the results of implementation of these tests on the data file are presented:

**Table 1: Results of ENT tests implementation.**

| Test Name | Result |
|---|---|
| **Entropy** | 7.999888 bits per byte |
| **Chi-square Test** | 252.64 for 1633342 samples, randomly exceed this value 53% of times |
| **Arithmetic Mean** | 127.4651 |
| **Monte Carlo Value For Pi** | 3.140154916 |
| **Serial Correlation Coeffcient** | 0.000019 |

Detailed description of tests and results is given below. Results of these tests confirm that gathered data consists of random numbers.

*Entropy:* The result of this test represents information density in bits per byte. Since there are eight bits in one byte, expected result is the number, which is close to eight. Obtained value from Table 1 confirms that number sequence from data file is random.

*Chi-square Test:* It is one of the most popular tests. Its result is represented as a real number with percentage which shows how frequently a random sequence would exceed this number. For truly random dataset percentage is expected to be between 10 and 90%. From Table 1 we can see that obtained value is almost exactly between these extreme values.

*Arithmetic Mean:* This test calculates arithmetic mean of the bytes of data file (their sum divided by the file length). For random dataset expected value is about 127.5 deviating randomly. Obtained result of this test is close to the expected.

*Monte Carlo Value for Pi:* Each 6 bytes of data are used as 24-bits coordinates of a point on 2D plane within a square. If the coordinates of this random point falls within the radius of the inscribed circle, this point is counted. Value for Pi is then obtained from the percentage of the counted points. The more precise is the resultant value - more random is the data. The test is sensitive and tends to fail (gives Pi=4) even for very small deviations from non-randomness. Obtained result from Table 1 is close to the value of Pi showing that data is sufficiently random. Note that this result is also dependent on the data size.

*Serial Correlation Coefficient:* This test searches for the dependence of the byte in data file on the previous byte. Expected value for the random sequence should be close to zero. From Table 1 we can see that no noticeable correlation exists in the data.

*The "Birthday" Test*

The test was performed using the DIEHARD test battery [19] to check the randomness of the sample. The test battery includes a large number of tests but they are more specific in nature, so only the first one from the available tests was chosen as the most illustrative.

The program chooses m days in a "year" of n days and lists the spacing between the birthdays. The number of values j that occurred more than once in that list is expected to be asymptotically Poisson distributed with mean $m^{\frac{3}{4n}}$. The test uses $n = 2^{24}$, and $m = 2^9$. The test runs nine times. 24 bits are used for each date generation. First time bits 1-24 (then the following 25-48, etc. until the end of the file) are used, then 2-25 (skipping first one) and so on to bits 9-32. Each time a chi-square test was used to check the goodness of fit and to get a p-value. For poor random number generator most of the p-values are expected to be close to 1. The results can be seen in the Table 2.

**Table 2: Results of the DIEHARD birthday spacing test.**

| | Duplicate spacing | | | | | | | | | |
|---|---|---|---|---|---|---|---|---|---|---|
| Used Bits | 0 | 1 | 2 | 3 | 4 | 5 | 6 to infinity | Mean | χ2/n.d.f. | p-value |
| Expected | 67.668 | 135.335 | 135.335 | 90.224 | 45.112 | 18.045 | 8.282 | 2 | | |
| Bits 1-24 | 56 | 147 | 146 | 89 | 41 | 15 | 6 | 1.964 | 5.39/6 | 0.50532 |
| Bits 2-25 | 59 | 124 | 150 | 93 | 50 | 17 | 7 | 2.062 | 4.52/6 | 0.39369 |
| Bits 3-26 | 70 | 143 | 158 | 66 | 40 | 14 | 9 | 1.884 | 12.36/6 | 0.94564 |
| Bits 4-27 | 69 | 159 | 130 | 74 | 44 | 17 | 7 | 1.892 | 7.58/6 | 0.72933 |
| Bits 5-28 | 79 | 136 | 126 | 89 | 45 | 20 | 5 | 1.936 | 4.07/6 | 0.33337 |
| Bits 6-29 | 59 | 121 | 138 | 107 | 47 | 23 | 5 | 2.102 | 8.54/6 | 0.798912 |
| Bits 7-30 | 60 | 137 | 151 | 83 | 43 | 19 | 7 | 2.002 | 3.63/6 | 0.273215 |
| Bits 8-31 | 65 | 142 | 134 | 95 | 40 | 18 | 6 | 1.964 | 1.91/6 | 0.071989 |
| Bits 9-31 | 70 | 130 | 130 | 105 | 40 | 15 | 10 | 2 | 4.37/6 | 0.373332 |

## Future Plans

The future plans include the building of the miniature functioning prototype of this device based on the small WSB microcontroller from hardkernel/Odroid-USBIO (http://www.hardkernel.com/main/products/prdt_info.php?g_code=G135390529643) to demonstrate the minimization and portability concept of the design.

## Conclusion

The results from the conducted tests show that the data collected using the presented design of the hardware random number generator is of high quality. Thus, the simple generator setup using the MPPC photosensor and the LED in conjuncture with the AMLS un-biasing algorithm presents a possibility for the future use in this capacity.

## Acknowledgments

The work was made possible by the support from NU School of Science and Technology, and NU grant #KF-14/19.